\documentclass{PoS}
\pdfoutput=1
\title{Spectra and elliptic flow of charmed hadrons in HYDJET++ model}

\ShortTitle{Spectra and elliptic flow of charmed hadrons in HYDJET++ model}

\author{\speaker{G. Eyyubova}$^{ab}$, I.P. Lokhtin$^a$, A. Belyaev$^a$, G. Ponimatkin$^c$ and E. Pronina$^a$\\
\llap{$^a$} Skobeltsyn Institute of Nuclear Physics, M.V. Lomonosov Moscow State University, Moscow, Russia\\
\llap{$^b$} Faculty of Nuclear Sciences and Physical Engineering, Czech Technical University in Prague,
Prague, Czech Republic\\
\llap{$^c$} Ostrov Industrial High School, Ostrov, Karlovy Vary District, Czech Republic\\
 E-mail: \email{Gyulnara.Eyyubova@fjfi.cvut.cz}
}

\abstract{Heavy-flavour quarks are predominantly produced in hard scatterings on a short time-scale and traverse the
medium interacting with its constituents, thus they are one of the effective probes of the transport properties of the medium formed in relativistic heavy ion collisions. On the other hand, the thermal production of heavy-flavour quarks in quark-gluon plasma (QGP) is itself of interest. 
In this report, the production and elliptic flow of the prompt charmed mesons $D^{0}$, $D^{+}$, $D^{∗+}$ and $J/\psi$ in PbPb collisions at the center-of-mass energy 2.76 TeV per nucleon pair are described in the framework of two-component HYDJET++ model. The model combines thermal and pQCD production mechanisms. The spectra and elliptic flow of charmed mesons are presented, the results are compared with LHC data.}

\FullConference{The European Physical Society Conference on High Energy Physics\\
		22--29 July 2015\\
		Vienna, Austria}

\begin{document}

\section{Introduction}
Studying the charmonium production is a particularly useful tool to probe the properties of hot and dense matter created in relativistic heavy ion collisions. In a picture, suggested by Matsui and Satz \cite{Matsui_Satz}, the production of charmonium states in QGP is suppressed due to the Debye screening of charm quarks and can provide the estimate of the system temperature. Later on, the enhancement scenarios via recombination of $c\bar{c}$ appeared as well \cite{Stat_coalesModel, Rafelski}. 
 
There are different mechanisms of charm production, contributing to the total yield. Generally, they can be divided into two main mechanisms: direct production in initial hard partonic scatterings and thermal production. 
The first one may be affected by a suppression and the second one can result in enhancement of charmonium yield. The thermal production itself can be both from the thermalized QGP (depending on the initial QGP temperature) and from the hadronic matter at hadronization stage. 
In order to favor different model assumptions more differential
studies on both hidden and open charm are required. While traversing the medium, heavy quarks interact with it and may participate in the collective expansion of the system and reach thermal equilibrium with the medium constituents. The azimuthal collective flow and nuclear modification factor carry the information on energy loss and degree of thermalization of heavy quarks in the medium.  

In this work, we study the $p_{\rm T}$-spectra and elliptic flow of charm hadrons (D, $J/\psi$) in PbPb collisions at $\sqrt{s_{NN}}$=2.76 TeV within HYDJET++ model, where the total charm yield has two origins. First, primordial D and $J/\psi$ mesons are produced in the initial hard scatterings from charm quark hadronization in vacuum (no color-screening effect, but taking into account medium-induced rescattering and energy loss of $b$- and $c$-quarks). Second, charmed mesons are produced at hadronization stage according to statistical coalescence approach. 

\section{HYDJET ++ model}
HYDJET++ \cite{hydjet} is the model of relativistic heavy ion collisions, which includes two independent components: the soft hydro-type state and the hard state based on pQCD with subsequent hadronization.

The simulation of the soft part is adapted from event generator FAST MC \cite{Amelin}. For the sake of simplicity and fast event generation there is no evolution from initial state, but only hadron production at freeze-out moment. Two freeze-outs with different temperatures are implemented in the model: the hadron multiplicities are defined at chemical freeze-out, while momentum distributions are defined at kinetic freeze-out. In order to simulate the elliptic flow effect, a spatial anisotropy of a thermal freeze-out volume is introduced, along with the azimuthal modulation of fluid velocities. The charm production (D, $J/\psi$ and $\Lambda_c$ hadrons) is treated within the statistical hadronization approach \cite{STAT_1, STAT_2}, where the yield of charmed hadrons $N_D$, $N_{J/\psi}$ is enhanced with respect to thermal yield $N^{th}_D$, $N^{th}_{J/\psi}$. The enhancement is characterized by charm enhancement factor (fugacity) $\gamma_c$. The $J/\psi$ yield is enhanced as:
$N_{J/\psi}=\gamma_c^2 N^{th}_{J/\psi}$,
while for open charmed hadrons the yield is enhanced as:
$\displaystyle N_{D}=\gamma_c N^{th}_{D}\frac{I_1(\gamma_c N^{th}_{D})}{I_0(\gamma_c N^{th}_{D})}$,
where $I_0$, $I_1$ are the modified Bessel functions.
The fugacity $\gamma_c$ can be treated as a free parameter of the model, or calculated through the number of charm quark pairs obtained from PYTHIA and multiplied by the number of NN sub-collisions.
 
The hard component uses PYTHIA \cite{PYTHIA} and PYQUEN \cite{PYQUEN} generators, which simulates parton energy loss in
the expanding quark-gluon fluid. High-momentum transfer limit \cite{HMTL_1,HMTL_2} is used for collisional loss. Radiative energy loss is based on BDMPS model \cite{BDMPS_1, BDMPS_2} and accounts for the "dead-cone" effect for massive quarks.
 
The input parameters of the model for soft and hard components have been tuned from fitting to heavy-ion data on various observables for inclusive hadrons at RHIC and LHC.
The parameter which regulates the contribution of each component to the total event is minimal $p_{\rm T}^{\rm min}$ of hard scattering. The partons produced in hard scatterings with transverse momentum transfer below $p_{\rm T}^{\rm min}$ do not contribute to hard part.
\section{Charmed meson $p_T$ spectra}
The charm enhancement factor $\gamma_c$ was tuned to describe the data on charmed meson yield at low $p_T$ in PbPb collisions at $\sqrt{s_{NN}} = 2.76$ TeV. Since the charmonuim yield is more sensitive to $\gamma_c$, the data on $J/\psi$ was used \cite{ALICE_jpsi} for this purpose. 
   
\begin{figure}
\includegraphics[width=0.5\textwidth]{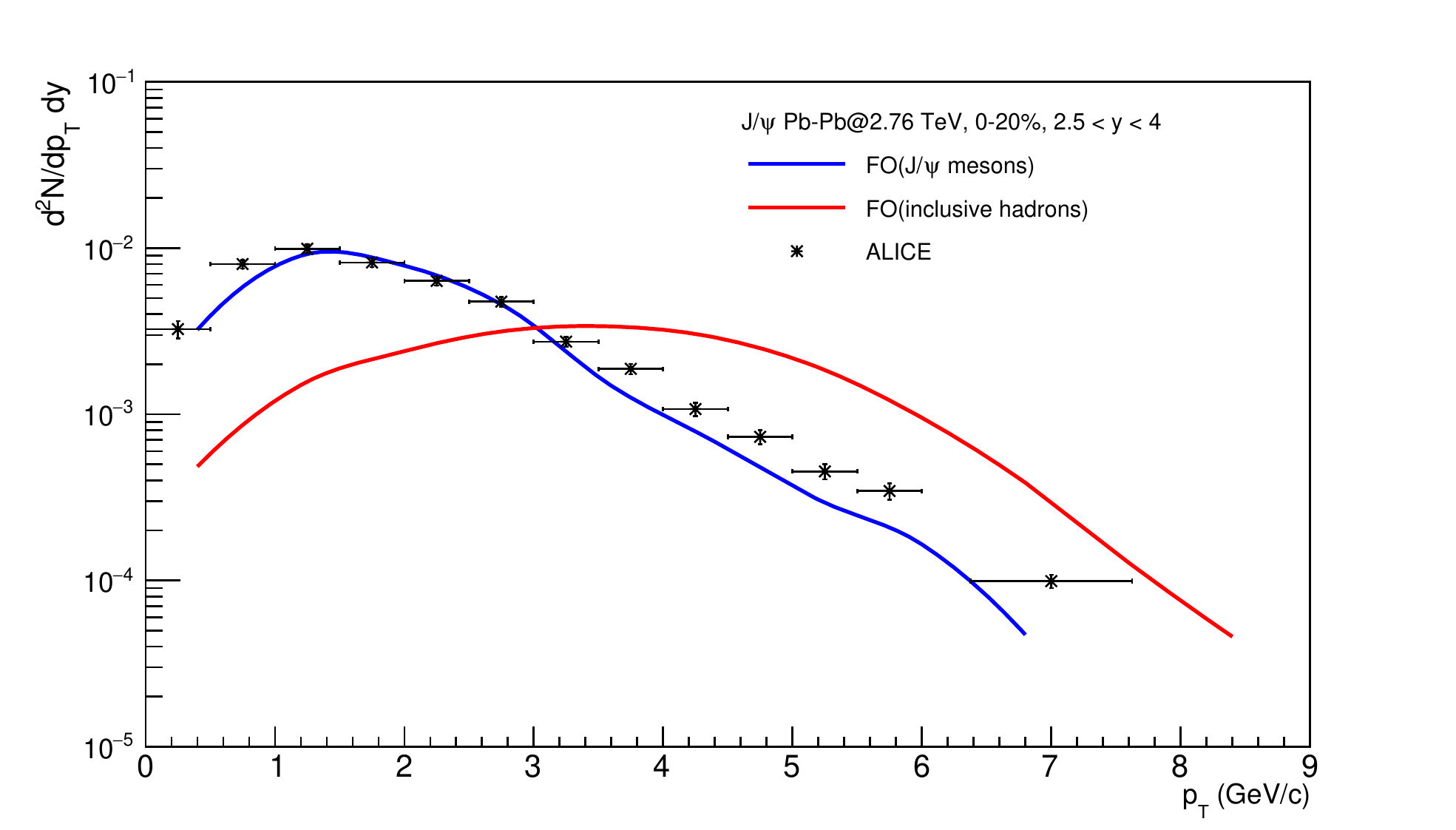}
\includegraphics[width=0.5\textwidth]{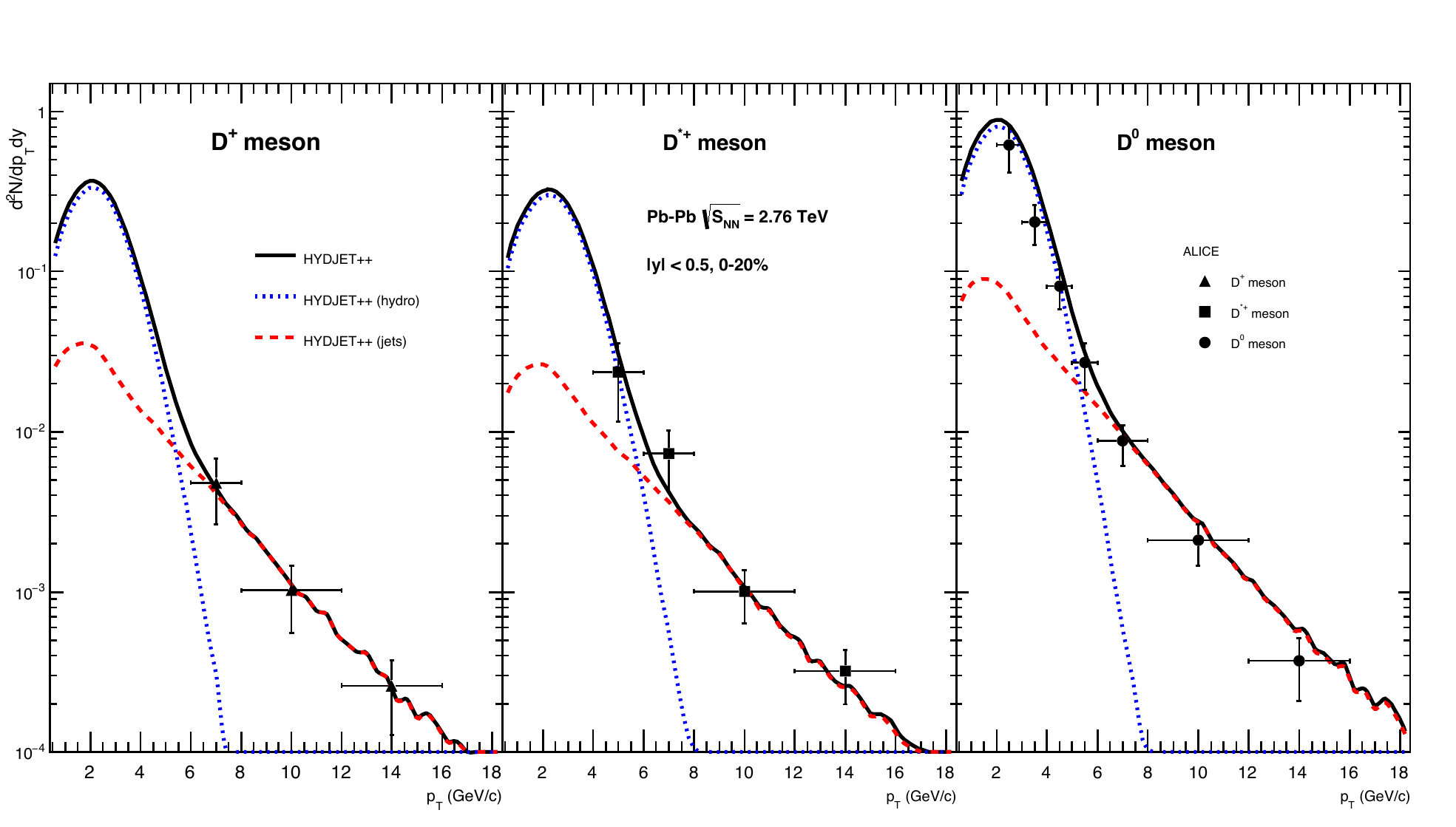}
\caption{Left: Transverse momentum spectrum of $J/\psi$-mesons at forward rapidity $2.5 < y < 4$ for 0-20\% centrality in PbPb collisions at $\sqrt{s_{NN}} = 2.76$ TeV. The points are ALICE data \cite{ALICE_jpsi},
histograms are simulated HYDJET++ events with two freeze-out scenarios: red line -- freeze-out parameters as for inclusive hadrons, blue line-- "early" thermal freeze-out. Right: Transverse momentum spectra of D-mesons at midrapidity $|y|< 0.5$ for 0-20\% centrality in PbPb collisions at $\sqrt{s_{NN}} = 2.76$ TeV. The points are ALICE data \cite{ALICE_dmeson}, histograms
are simulated HYDJET++ events (blue line -- soft component, red line -- hard component, black line -- sum of two components). }
\label{fig_jpsi}
\end{figure}

Previously in \cite{hydjet_charm} it was demonstrated that the data on $p_T$ and $y$ spectra of $J/\psi$-meson at RHIC energy $\sqrt{s_{NN}}=200$ GeV \cite{PHENIX_jpsi} can be reproduced by the HYDJET++ model with the assumption that the thermal freeze-out for $J/\psi$-mesons happens at the same temperature as chemical freeze-out. The early thermal freeze-out of $J/\psi$ was already suggested before in order to describe SPS data at 158 GeV/nucleon \cite{early_freeze-out}. It turned out that the similar situation holds at the LHC. Figure \ref{fig_jpsi}, left, shows the comparison of HYDJET++ simulations with the ALICE data \cite{ALICE_jpsi} for $p_T$ spectrum of $J/\psi$-mesons in 20\% of most central PbPb collisions at $\sqrt{s_{NN}} = 2.76$ TeV. One can see that if thermal freeze-out for $J/\psi$ happens at the same temperature as chemical freeze-out (with reduced collective velocities), then simulated spectrum matches the data up to $p_T \sim 3$ GeV/c. The contribution of hard component had to be increased for better description of the shape of $p_T$-spectrum. At the same time, the data on open charmed mesons can be reproduced using the same freeze-out parameters and hard component contribution as for all inclusive hadrons. Figure \ref{fig_jpsi}, right, shows the comparison of HYDJET++ simulations with the ALICE data \cite{ALICE_dmeson} for $p_T$-spectra of $D^0$, $D^{\pm}$, $D^{*\pm}$ in 20\% of most central PbPb collisions at $\sqrt{s_{NN}} = 2.76$ TeV. The fugacity value $\gamma_c$ = 11.5 is fixed from $J/\psi$ yield.

\section{Charmed meson elliptic flow}
\begin{figure}
\includegraphics[width=0.5\textwidth]{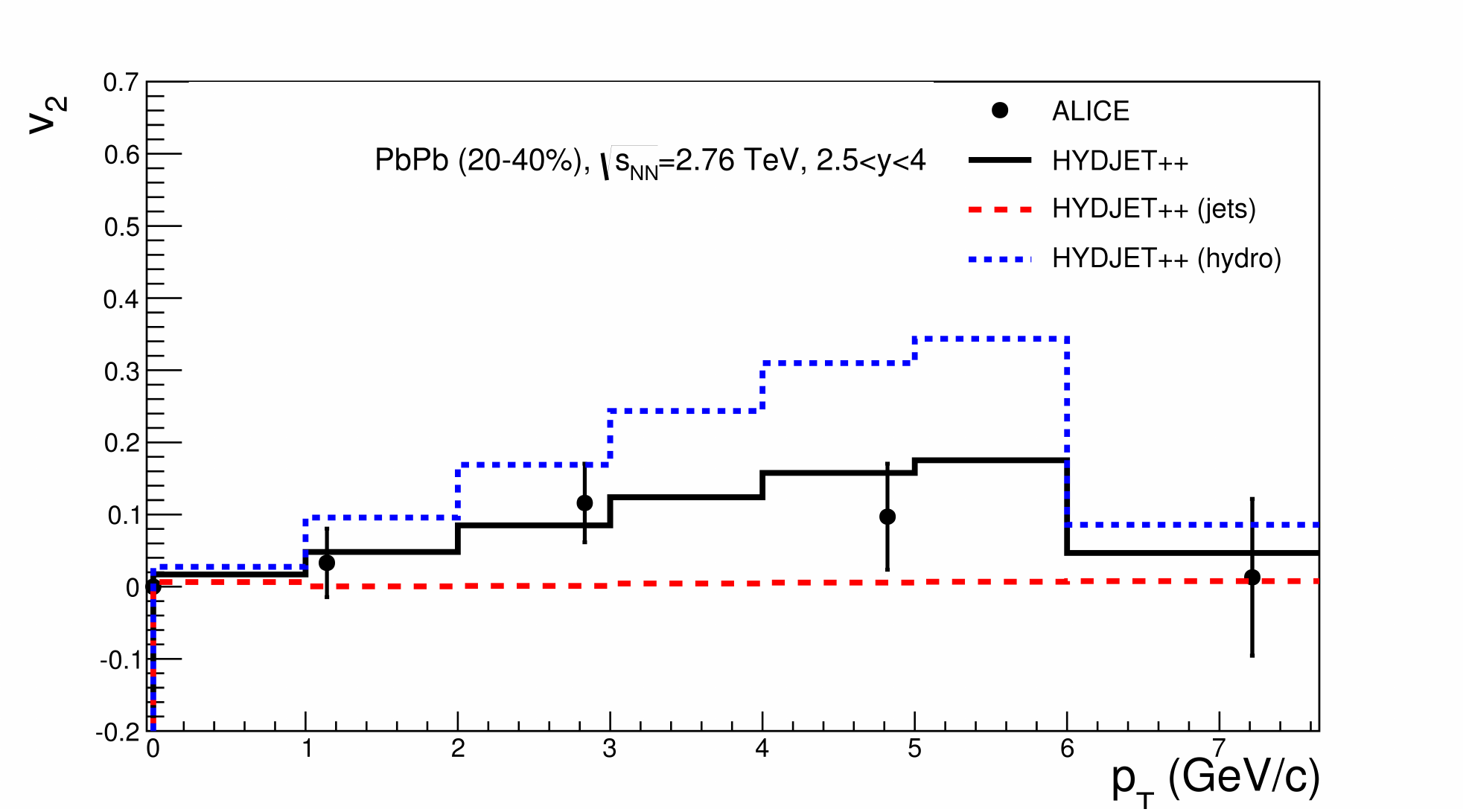}
\includegraphics[width=0.5\textwidth]{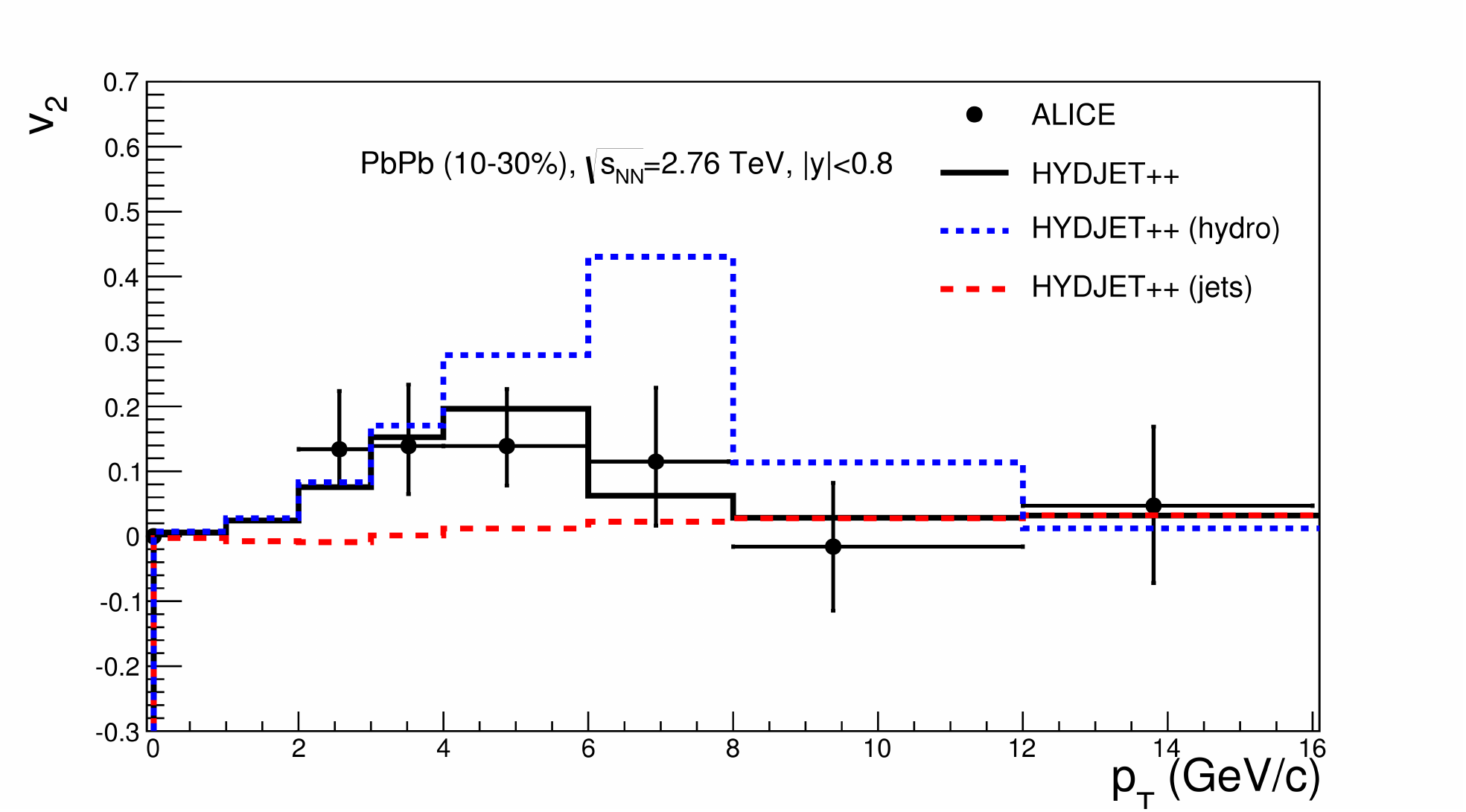}
\caption{Left: Elliptic flow $v_2(p_T)$ of $J/\psi$-mesons at forward rapidity $2.5 < y < 4$ for 20-40\% centrality of PbPb collisions at $\sqrt{s_{NN}} = 2.76$ TeV. The points are ALICE data \cite{ALICE_jpsiv2},histograms are simulated HYDJET++ events (blue line --soft component, red line -- hard component, black line -- sum of two components).
Right: Elliptic flow $v_2(p_T)$ of $D^0$-mesons at midrapidity $|y|< 0.8$ for 10-30\% centrality of PbPb collisions at $\sqrt{s_{NN}} = 2.76$ TeV. The points are ALICE data \cite{ALICE_dmesonv2}, histograms
are simulated HYDJET++ events (blue line -- soft component, red line -- hard component, black line -- sum of two components). }
\label{fig_jpsiv2}
\end{figure}

The elliptic flow coefficient $v_2$ characterizes the azimuthal asymmetry of the yield with respect to event plane angle. The positive elliptic flow of D-mesons and $J/\psi$, measured experimentally \cite{ALICE_jpsiv2, ALICE_dmesonv2}, indicates that heavy charm quarks inherit a collective azimuthal flow and thus may be partially in thermal equilibrium with the medium. 
Figure \ref{fig_jpsiv2} shows the comparison of HYDJET++ model results with the ALICE data \cite{ALICE_jpsiv2, ALICE_dmesonv2} for $p_T$ dependence of elliptic flow $v_2(p_T)$ of $J/\psi$-mesons for 20-40\% centrality (left) and D-mesons for 10-30\% centrality (right) of PbPb collisions at $\sqrt{s_{NN}} = 2.76$ TeV. The parameters of the model and freeze-out conditions are the same, as for the spectra description.   
Within the experimental uncertainties the model results agree with the data. 
\section{Conclusion}
Momentum spectra and elliptic flow of D and $J/\psi$ mesons in PbPb collisions at the LHC are reproduced by two-component model HYDJET++, which combines thermal and non-thermal charm production. 
The model configuration for soft and hard part, used to describe inclusive hadron production, allows to reproduce the data on D-meson reasonably well. Thus, the significant part of D-mesons (up to $p_T \sim$ 4 GeV/c) seems to
be in a kinetic equilibrium with the medium. 

In order to describe data on $J/\psi$-meson the scenario of early thermal freeze-out was used, besides, the contribution of hard part was increased. Thus, the thermal freeze-out of $J/\psi$-mesons happens appreciably before freeze-out of light hadrons, presumably at chemical freeze-out, which means that the significant part of $J/\psi$-mesons are not in a kinetic equilibrium with the medium.

\section{Acknowledgments}
Discussions with L.V. Bravina, V.L. Korotkikh, L.V. Malinina, A.M. Snigirev, E.E. Zabrodin and S.V. Petrushanko are gratefully acknowledged. This work was supported by the Russian Scientific Fund under Grant No. 14-12-00110 in a part of computer simulation of $p_T$-spectrum and elliptic flow
of charmed mesons in PbPb collisions. G.E. acknowledges the European Social Fund within the framework of realizing the project Support of Inter-sectoral Mobility and Quality Enhancement of Research Teams at Czech Technical University in Prague, CZ.1.07/2.3.00/30.0034.

\end{document}